\documentclass[11pt,a4paper]{article}
\usepackage{amsmath,amsfonts,epsfig,color,latexsym}
\usepackage{amssymb}
\usepackage[english, USenglish]{babel}
\usepackage{epsfig,psfrag}
\usepackage{a4wide}
\usepackage{rotating}

\usepackage{hyperref}

\csname @addtoreset\endcsname{equation}{section}

\makeatletter
\def\timenow{\@tempcnta\time
  \@tempcntb\@tempcnta
  \divide\@tempcntb60
  \ifnum10>\@tempcntb0\fi\number\@tempcntb
  \multiply\@tempcntb60
  \advance\@tempcnta-\@tempcntb
  :\ifnum10>\@tempcnta0\fi\number\@tempcnta}
\makeatother

\begin{document}

\renewcommand{\thefootnote}{\fnsymbol{footnote}}
\newpage
\pagestyle{empty}
\setcounter{page}{0}


\newcommand{\norm}[1]{{\protect\normalsize{#1}}}
\newcommand{\p}[1]{(\ref{#1})}
\newcommand{\half}{{\ts \frac{1}{2}}}
\newcommand \vev [1] {\langle{#1}\rangle}
\newcommand \ket [1] {|{#1}\rangle}
\newcommand \bra [1] {\langle {#1}|}

\newcommand{\cM}{{\cal M}} 
\newcommand{\cR}{{\cal R}} 
\newcommand{\cS}{{\cal S}} 
\newcommand{\cK}{{\cal K}}
\newcommand{\cL}{{\cal L}} 
\newcommand{\cF}{{\cal F}}
\newcommand{\cN}{{\cal N}}
\newcommand{\cA}{{\cal A}}
\newcommand{\cB}{{\cal B}}
\newcommand{\cG}{{\cal G}}
\newcommand{\cO}{{\cal O}}
\newcommand{\cY}{{\cal Y}}
\newcommand{\cX}{{\cal X}}
\newcommand{\cT}{{\cal T}}
\newcommand{\cW}{{\cal W}}
\newcommand{\cP}{{\cal P}}
\newcommand{\nt}{\notag\\} 
\newcommand{\pa}{\partial}
\newcommand{\ep}{\epsilon}
\newcommand{\om}{\omega}
\newcommand{\bep}{\bar\epsilon}
\renewcommand{\a}{\alpha}
\renewcommand{\b}{\beta}
\newcommand{\g}{\gamma}
\newcommand{\s}{\sigma}
\newcommand{\la}{\lambda}
\newcommand{\da}{{\dot\alpha}}
\newcommand{\db}{{\dot\beta}}
\newcommand{\dg}{{\dot\gamma}}
\newcommand{\dd}{{\dot\delta}}
\newcommand{\q}{\theta}
\newcommand{\bq}{\bar\theta}
\newcommand{\bQ}{\bar Q}
\newcommand{\tx}{\tilde{x}}
\newcommand{\tr}{\mbox{tr}}

\vspace{20mm}

\bigskip
\bigskip

\null\vskip-53pt \hfill
\begin{minipage}[t]{50mm}
CERN-TH-2016-071 \\
LAPTH-018/16
\end{minipage}

\vskip1.1truecm
\begin{center}
\vskip 0.0truecm

 {\Large\bf
Demystifying the twistor construction of composite operators in $\cN=4$ super-Yang-Mills theory}
\vskip 1.0truecm

{\bf  Dmitry Chicherin$^{a}$ and  Emery Sokatchev$^{a,b}$ \\
}

\vskip 0.4truecm
 $^{a}$ {\it LAPTH\,\footnote{Laboratoire d'Annecy-le-Vieux de Physique Th\'{e}orique, UMR 5108},   Universit\'{e} de Savoie, CNRS,
B.P. 110,  F-74941 Annecy-le-Vieux, France\\
  \vskip .2truecm

$^{b}$ Theoretical Physics Department, CERN, CH -1211, Geneva 23, Switzerland 
} \\
\end{center}

\vskip .3truecm

\centerline{\bf Abstract} 
\medskip
\noindent
We explain some details of the construction of composite 
operators in $\cN=4$ SYM that we have elaborated earlier in the context of Lorentz harmonic chiral (LHC) superspace. We give a step-by-step elementary derivation and show that the result coincides with the recent hypothesis put forward in arXiv:1603.04471 within the twistor approach. We provide the appropriate LHC-to-twistors dictionary.
\newpage

\newpage\setcounter{page}{1}\setcounter{footnote}{1}
\pagestyle{plain}
\renewcommand{\thefootnote}{\arabic{footnote}}

\section{Introduction}

In this note we explain some details of the construction of composite 
operators in $\cN=4$ SYM elaborated in \cite{Chicherin:2016fac} in the context of Lorentz harmonic chiral (LHC) superspace. 
There we gave a rigorous {\it proof} of the construction using Lorentz harmonics. The harmonic formalism \cite{Galperin:1984av,Galperin:2001uw} is essentially equivalent \cite{Galperin:1987wc,Evans:1992az,Lovelace:2010ev} to the twistor approach, although there are some conceptual differences.
The twistor construction of composite operators has been pioneered in \cite{Adamo:2011dq,Adamo:2011cd} in the aim of proving the duality between correlators on the light cone and light-like polygonal Wilson loops \cite{Alday:2010zy,Eden:2010zz}.
However, the construction presented there is incomplete and is based on a `natural guess'. 
The first complete treatment of a composite operator in the twistor framework has been given in \cite{Chicherin:2014uca} for 
the chiral truncation of the stress-tensor multiplet, but it applies only to that specific case. A general construction valid for all kinds of composite operators has been developed in \cite{Chicherin:2016fac} using the conceptually simpler LHC approach.
The authors of the recent note \cite{Koster:2016ebi} have applied the twistor construction of composite operators from \cite{Adamo:2011dq}  to the 
calculation of MHV form-factors initiated earlier in \cite{Koster:2014fva}. They have realized that the existing twistor construction does not lead to the correct result. 
In order to reproduce the known form-factors they put forward a {\it hypothesis}
how to adjust the twistor construction of scalar composite operators (the bottom components of, e.g., the Konishi  supermultiplet and others). Here would like to emphasize that our LHC construction proposed earlier in Ref.~\cite{Chicherin:2016fac} not only gives the same result but it also provides a step-by-step derivation of the twistor-based hypothesis of \cite{Koster:2016ebi}. 
We show that the twistor formulae from \cite{Koster:2016ebi} coincide with the harmonic formulae from \cite{Chicherin:2016fac} and we provide the appropriate LHC-to-twistors dictionary. 

We hope that this note will help to build a linguistic bridge between the twistor and harmonic communities.

\section{$\cN = 4$ SYM in Lorentz harmonic chiral superspace}

In Ref.~\cite{Chicherin:2016fac} the $\cN = 4$ SYM theory has been reformulated in terms of Lorentz harmonic analytic superfields.
This off-shell formulation is closely related to the twistor approach of L. Mason et al \cite{Mason:2005zm,Boels:2006ir,Adamo:2011cb}.  
It has the chiral ($Q$) supersymmetry and the $SU(4)$ R-symmetry of the $\cN = 4$ SYM theory manifest.
The price to pay for having half of the supersymmetry explicit off shell is the infinite number of auxiliary fields and pure gauges of arbitrarily high spin. 
In order to handle them conveniently we introduce 
auxiliary variables, the so-called Lorentz harmonics~\cite{Galperin:1984av,Galperin:2001uw,Roslyi:1989ya,Roslyi:1986yg,Devchand:1992st,Devchand:1993ba,Devchand:1995gp}.  This situation is not new, it is very reminiscent of the off-shell formulation of the $\cN=2$ hypermultiplet in harmonic superspace with an infinite set of auxiliary fields \cite{Galperin:1984av,Roslyi:1989ya}. 

We work in Euclidean four-dimensional space with Lorentz group $SO(4) \sim SU(2)_{L} \times SU(2)_{R}$. The left and right factors act on the undotted and dotted Lorentz indices of the space-time coordinates $x^{\da \a}= x^\mu\tilde\sigma^{\da\a}_\mu $, respectively.
The harmonic variables $u^{+}_{\a}$ and $u^-_{\a}$ are a pair of $SU(2)_L$ spinors forming an $SU(2)_L$ matrix:
\begin{equation}\label{2}
\parallel u\parallel\  = \left(\begin{array}{cc} u^+_1 & u^-_1 \\
u^+_2 & u^-_2 \end{array} \right) \in SU(2)_L\ .  
\end{equation}
This implies the normalization condition $u^{+\a}u^-_{\a} = 1$ and the complex conjugation rules $(u^+_{\a})^* = - u^{-\a}$ 
and $(u^{+\a})^* = u^{-}_{\a}$ (the $SU(2)$ indices are raised and lowered with the two-dimensional Levi-Civita tensor, $u^\a=\ep^{\a\b} u_\b$ with $\ep^{12}=-1$). The harmonics provide a global description of the compact coset $S^2 \sim SU(2)_L/U(1)$, so that their indices $\pm$ refer to the $U(1)$ charge.
In what follows we deal with nonlocal expressions in harmonic space depending on multiple sets of harmonics (copies of $S^2$).
The $U(1)$ symmetry is local, i.e. all the expressions are covariant with respect to the $U(1)$ transformations of the harmonics on each copy of $S^2$. At the same time $SU(2)_L$ is a global symmetry, i.e. it acts simultaneously on the indices $\a,\b,\ldots$  of all the harmonics.

The (super)fields we are going to use are harmonic functions.
They are defined by their harmonic expansion on $S^2$,
\begin{equation}\label{3}
f^{(q)}(u) = \sum^\infty_{n=0} f^{\a_1 \ldots \a_{2n+q}}
u^+_{(\a_1} \ldots u^+_{\a_{n+q}} u^-_{\a_{n+q+1}}
\ldots u^-_{\a_{2n+q})} \qquad \text{for} \quad q \geq 0\,,
\end{equation}
and carry a definite charge (degree of homogeneity) $q$.  Each $u-$monomial is a spherical harmonic of spin $n+q/2$ in a coordinateless realization. From (\ref{3})
it is clear that the harmonic functions are collections of infinitely many finite-dimensional 
irreducible representations of $SU(2)_L$ (totally symmetric multispinors). 
So, a harmonic field $f^{(q)}(x,u)$ consist of an 
infinite number of ordinary multispinor fields $f^{\a_1 \ldots \a_{m}}(x)$.

The differential operators compatible with the normalization condition $u^{+\a}u^-_{\a} = 1$ 
are the covariant harmonic derivatives
\begin{align}\label{4}
\pa^{++}= u^{+\a}{\pa\over\pa u^{-\a}} \ , \ \ \
\pa^{--}= u^{-\a}{\pa\over\pa u^{+\a}}\,,
\end{align}
having the meaning of the raising and lowering operators of $SU(2)_L$,  and 
the Cartan charge $\pa^0$ of $SU(2)_L$ which counts the $U(1)$ charge of the harmonic functions, $\pa^0f^{(q)}(u) = q f^{(q)}(u)$. These three derivatives form the algebra of $SU(2)_L$,
\begin{align}\label{su2}
[\pa^{++}, \pa^{--}] = \pa^0 \;,\qquad [\pa^0, \pa^{++}] = 2 \pa^{++} \;,\qquad [\pa^0, \pa^{--}] = -2 \pa^{--} \,.
\end{align} 
The restriction to functions on $SU(2)_L$ with definite charge gives a particular realization of the harmonic coset $SU(2)_L/U(1)$. An important property of the harmonic functions of zero charge is that they become constants  if they satisfy the constraint 
\begin{align}\label{Han}
\pa^{++}f^{(0)}(u)=0\quad \Rightarrow \quad   f^{(0)}(u)=\rm const.
\end{align}
It follows directly from the expansion \p{3} or from the fact that \p{Han} defines an $SU(2)_L$ highest weight of charge zero, hence a singlet. 

We also need an $SU(2)_L$ invariant harmonic integral on $S^2$. 
It amounts to extracting the singlet part of
a chargeless integrand,  according to the  rule
\begin{equation}\label{6}
\int du\; f^{(q)}(u) =  \left\{ \begin{array}{ll}
0,  \ q \neq 0 \\ f_{\rm singlet} ,  \ q= 0 \end{array} \right.\ .
\end{equation}
In particular, $\int d u = 1$. 
This rule is compatible with integration by parts for the harmonic
derivatives \p{4}.

The harmonics allow us to decompose, in a Lorentz covariant way, the chiral odd superspace variables $\q^{\a A}$ (with $A =1 , \ldots,4$  being an $SU(4)$ R-symmetry index) into a pair of odd variables $\q^{+ A}$ and $\q^{-A}$,
\begin{align}
\q^{\a A} = u^{+\a} \q^{- A} + u^{-\a} \q^{+ A}\ \ \ , \ \ \ \q^{\pm A} \equiv u^{\pm}_{\a} \q^{\a A}\,.
\end{align}
We use superfields of two types: chiral harmonic superfields $\Phi(x^{\da\a},\q^{\a A},u)$
and Lorentz-analytic (L-analytic) chiral superfields $\Phi(x^{\da \a},\q^{+A},u)$.
L-analyticity is a kind of Grassmann analyticity and it means that the superfields do not depend on $\q^{-A}$. 
L-analytic superfields are covariant only with respect to the chiral $Q$-half of supersymmetry, but the antichiral 
$\bQ$-supersymmetry is not manifest. 
By abuse of language we call them superfields although, strictly speaking, they are semi-superfields.

Alongside with the harmonic derivatives we define harmonic projected odd and space-time derivatives,
\begin{align}\label{dif}
\pa^\pm_A \equiv u^{\pm \a} \frac{\pa}{\pa \q^{\a A}}\ \ \ , \ \ \ 
\pa^{\pm}_\da \equiv u^{\pm \a} \frac{\pa}{\pa x^{\da\a}}\,.
\end{align}
An L-analytic superfield $\Phi(x^{\da \a},\q^{+A},u)$ satisfies the analyticity condition $\pa^+_{A} \Phi=0$.
In order to formulate the $\cN = 4$ SYM theory we consider gauge transformations whose parameter $\Lambda(x,\q^+,u)$ is an 
L-analytic chiral superfield of vanishing $U(1)$ charge.
Then all the derivatives except for $\pa^0$ and $\pa^+_{A}$ (because $\pa^0 \Lambda = \pa^+_{A} \Lambda = 0$) acquire gauge connections, $\pa \to \nabla=\pa+A$:
\begin{align}\label{covar}
\pa^{++} \to \nabla^{++} \ , \ \pa^{--} \to \nabla^{--} \ , \ \pa^0 \to \pa^0 \ , \
\pa^+_A \to \pa^+_A  \ , \ \pa^-_A \to \nabla^-_A \ , \ \pa^\pm_{\da} \to \nabla^{\pm}_{\da}\,.
\end{align}
The infinitesimal gauge transformations of the connections $A$ have the usual form
\begin{equation}\label{21} \delta_{\Lambda} A = \pa \Lambda + [A, \Lambda] 
= \nabla \Lambda \;\;, \qquad \Lambda=\Lambda(x,\q^+,u) \,.
\end{equation}
The gauge connections $A^{++}$ and $A^+_\da$ for the derivatives $\pa^{++}$ and $\pa^+_{\da}$ 
are the main objects of the $\cN=4$ SYM theory, or the {\it gauge prepotentials}. They are L-analytic chiral superfields,
\begin{align}\label{3.3}
\nabla^{++} = \pa^{++} + A^{++}(x,\q^+,u)   \,, \qquad \nabla^+_\da =  \pa^+_\da + A^+_\da(x,\q^+,u) \,.
\end{align}
The remaining gauge connections $A^{--}$, $A^-_{A}$ and $A^{-}_{\da}$ (we will need only the first two)
can be expressed in terms of $A^{++}$ and $A^+_{\da}$. They are chiral but not L-analytic (they depend on all the eight chiral odd variables).

The connections $A^{--}$, $A^-_{A}$ are obtained by covariantizing the commutation relations 
$[\pa^{++},\pa^{--}] = \pa^0$ and $[\pa^{--},\pa^+_{A}] = \pa^-_{A}$ (see eqs.~\p{4}, \p{su2} and \p{dif}),
\begin{align}\label{covarr}
[\nabla^{++}, \nabla^{--}] = \pa^0 \ \ , \ \  [\nabla^{--}, \pa^+_A] = \nabla^-_A \,.
\end{align}
The first  relation in \p{covarr} is
a harmonic differential equation for the unknown harmonic connection $A^{--}$ in terms of the given $A^{++}$.
It has a unique solution which is non-polynomial in $A^{++}$ \cite{Zupnik:1987vm,Galperin:2001uw},
\begin{align}\label{448}
A^{--}(x,\q,u) =  \sum^\infty_{n=1} (-1)^n\int du_1\ldots du_n\; { 
A^{++}(x,\q \cdot u^+_1,u_1) \ldots A^{++}(x,\q \cdot u^+_n,u_n) \over (u^+u^+_1)(u^+_1u^+_2) \ldots
(u^+_nu^+)}
\end{align} 
with $\q^{A} \cdot u^+_k = \q^{\a A} (u^+_{k})_{\a}$.
Eq.~\p{448} is local in $(x,\q)$ space but nonlocal in harmonic space. Once we have found 
$A^{--}$, the remaining gauge connections and all the (super)curvatures can be directly expressed in its terms. For example, the second relation in \p{covarr} enables us to find the gauge connection $A^{-}_{A}$,
\begin{align}\label{3.9}
A^-_A = - \pa^+_A A^{--}\,.
\end{align}

The familiar degrees of freedom of the on-shell $\cN = 4$ vector multiplet are obtained after eliminating an infinite number of 
auxiliary fields and pure gauges contained in the harmonic expansions of the prepotentials $A^{++}$ and $A^+_{\da}$.
By means of a gauge transformation \p{21} we can fix the Wess-Zumino gauge
\begin{align}\label{23}
A^{++}&=\frac{1}{2} \q^{+A}\q^{+B} \phi_{AB}(x) + \frac{1}{6} \ep_{ABCD} \q^{+B}\q^{+C}\q^{+D} u^{-\a}\psi^A_{\a}(x) \nt
&+ \frac{1}{8} \ep_{ABCD} \q^{+A}\q^{+B}\q^{+C}\q^{+D} u^{-\a}u^{-\b} G_{\a\b}(x)\,.
\end{align}
It leaves only the 6 scalars  $\phi_{AB}$, the 4 chiral gluinos $\psi^{A}_{\a}$ and the self-dual Lagrange multiplier $G_{\a\b}=G_{\b\a}$. 

Further, the infinite set of auxiliary fields contained in $A^{+}_{\da}$ is eliminated by imposing  constraints 
which are equivalent to equations of motion. In \cite{Chicherin:2016fac} we identified the relevant constraints as
\begin{align}\label{3.14'}
\pa^+_\da A^{++} - \pa^{++} A^+_\da + [A^+_\da,
A^{++}] = 0\;\; , \qquad
\pa^{+\da} A^+_\da + A^{+\da} A^+_\da = (\pa^+)^4 A^{--}\,.
\end{align}
 By partially solving the first of them the field content of $A^+_{\da}$ is reduced to the gluon $\cA_{\a\da}$ and the four antichiral gluions $\bar\psi_{\da A}$,
\begin{align}\label{25'}
A^{+}_\da = u^{+\a}{\cal A}_{\a\da}(x) + \q^{+A}
\bar\psi_{\da A}(x) + \text{derivative terms}.
\end{align}
In addition, the first equation in \p{3.14'} 
imposes the constraint $\nabla^{\da\a} G_{\a\b} = 0$ (with $\nabla^{\da\a} = \pa^{\da\a} + \cA^{\da\a}$) and the second equation implies $F_{\a\b} = G_{\a\b}$,
where $F_{\a\b}$ is the self-dual part of the YM curvature $F_{\mu \nu}$.
Let us emphasize once more 
that in order to identify the familiar gluon, gluinos and scalars we have to first eliminate 
all the auxiliary fields and pure gauges. This automatically puts the remaining physical fields on shell. 
For completeness we also show the component content of $A^{--}$, eq.~\p{448} (Abelian or free case):
\begin{align}\label{512}
& A^{--} = \frac{1}{2} \q^{-A}\q^{-B} \phi_{AB}
+ \frac1{4}\ep_{ABCD}\q^{-B} \q^{-C} \Bigl(\q^{+D} u^-_{\a} - \frac1{3} u^{+}_{\a} \psi^{+A}\Bigr)\psi^{\a A} \nt
& \qquad  + \frac{1}{6}\ep_{ABCD} \q^{-C}\q^{-D}\Bigl(\frac{1}{12}\q^{-A}\q^{-B} u^+_{\a} u^+_{\b} 
- \frac{1}{3}\q^{+A}\q^{-B} u^+_{\a} u^-_{\b} 
+  \frac{1}{2}\q^{+A}\q^{+B} u^-_{\a} u^-_{\b} \Bigr) G^{\a\b} + O(g)\,.
\end{align}

The equations of motion \p{3.14'} can be derived from the following off-shell action \cite{Chicherin:2016fac}:
\begin{equation} \label{N4}
S_{\cN=4\ SYM} = \int d^4x du d^4\q^+\; L_{\rm CS}(x,\q^+,u) + \int d^4x d^8\q\; L_{\rm Z}(x, \q) \,.
\end{equation} 
Here $L_{\rm CS}$ is a Chern-Simons-like Lagrangian, which is L-analytic and involves both prepotentials  $A^{++}$ and $A^{+}_{\da}$. It describes the self-dual sector of the theory \cite{Siegel:1992xp,Sokatchev:1995nj,Witten:2003nn}. The second term $L_{\rm Z}$ is chiral, but not L-analytic. It involves only the prepotential $A^{++}$ and has a form coinciding with the $\cN=2$ SYM action in harmonic superspace as given by Zupnik \cite{Zupnik:1987vm}. The role of this term is to complete the self-dual sector to the full super-Yang-Mills theory \cite{Mason:2005zm}. The pure YM truncation of the action \p{N4} is a first-order formulation using the Lagrange multiplier $G_{\a\b}$ for the YM curvature $F_{\a\b}$ \cite{Chalmers:1997sg}.

Concluding this brief overview of the LHC formulation of $\cN=4$ SYM, a few words about the equivalent twistor approach. In it the  harmonics $u^+_\a$ are replaced by holomorphic spinor coordinates $\lambda_{\a}$ on $\mathbb{CP}^1$ fibers.
The harmonic derivative $\pa^{++}$ corresponds to the twistor derivative $\bar\pa|_{x,\q}$.
The local $U(1)$ charge of the harmonic functions \p{3} corresponds to the degree of  homogeneity of the functions on the (super)twistor space $\mathbb{CP}^{3|4}$
with coordinates $\mathcal{Z} = \lambda_{\b}(\ep^{\b\a} ,i x^{\da\b}, i \q^{\b A})$. 
The gauge connections $A^{++}$ and $A^{+}_{\da}$ are replaced by a (0,1)-form $\cA$ which 
lives on $\mathbb{CP}^{3|4}$ \cite{Adamo:2011cb}. The L-analyticity of $A^{++}$ and $A^{+}_{\da}$ implies that the (0,1) form $\cA$ depends on the 
holomorphic projection $\chi^A = \lambda_{\a} \q^{\a A}$ of the odd variable $\q$ but is independent of $\hat\lambda_{\a}\q^{\a A}$, 
where $\hat\lambda$ is the Euclidean conjugate of $\lambda$. So, $\la$ and $\hat\lambda$ are the analogs of the harmonics $u^+_\a$ and $u^-_\a$, respectively. The physical fields are extracted from $\cA$ by an integral  Penrose transform \cite{Penrose:1986ca},
that corresponds to picking out the first term in the harmonic expansion \p{3}.
The harmonic measure $d u \sim (u^{+\a} du^+_\a) (u^{-\b} du^-_\b)$ carries zero $U(1)$ charge, while in the twistor approach the projective measure 
$D \lambda = \langle \lambda d \lambda \rangle$ (the equivalent of $u^{+\a} du^+_\a$) has degree of  homogeneity $2$. The simple integration rule \p{6}, based on $SU(2)_L$ covariance, is replaced by contour integration on $\mathbb{CP}^1$.

The main conceptual difference between the twistor and harmonic approaches 
is the absence of a twistor analog of the harmonic gauge connection $A^{--}$ and of the associated  $SU(2)_L$
algebraic structure \p{su2}. The notion of $A^{--}$ is very useful and makes the construction of composite operators in the next section straightforward and very easy.

\section{Composite operators} \label{CompOp}

The standard formulation of the on-shell $\cN = 4$ vector multiplet  uses the supercurvature $W_{AB}(x,\q,\bq)= -W_{BA}$
appearing in the defining constraints \cite{Sohnius:1978wk}
\begin{align}\label{constr}
\{\nabla^{\a}_A , \nabla^{\b}_B\} = \ep^{\a\b} W_{AB} \ \ , \ \ \{\bar\nabla_{\da}^A , \bar\nabla_{\db}^B\} = \ep_{\da\db}\bar W^{AB} \ \ , \ \
\{\nabla_{\a A} , \bar\nabla^{B}_{\da}\} = \delta^B_A \nabla_{\a\da}
\end{align}
and satisfying the reality condition $W_{AB} = \frac{1}{2}\ep_{ABCD} \bar W^{CD}$. These constraints imply the equations of motion. 
Given the supercurvature $W_{AB}(x,\q,\bq)$ one can construct all the gauge-invariant local operators in the theory. For example, among the simplest bilinear operators of twist two one finds 
the Konishi multiplet 
\begin{align}\label{K}
K(x,\q,\bq) = \tr\, (W_{AB} \bar W^{AB})= \frac{1}{2}\ep^{ABCD} \tr (W_{AB} W_{CD})\,,
\end{align}
and the protected half-BPS multiplet in the ${\mathbf {20'}}$ of $SU(4)$
\begin{align}\label{}
O_{\mathbf {20'}}  = \tr\, (W_{AB} W_{CD}) -\frac1{12} \ep_{ABCD}\, \tr\, (W_{EF} \bar W^{EF})\,.
\end{align}

The LHC (or the equivalent twistor) formulation makes only the chiral half of supersymmetry manifest. The chiral truncation (obtained by setting $\bq=0$) of the supercurvature $W_{AB}(x,\q)$ appears in the covariantized 
anticommutation relation $\{ \pa^-_A , \pa^+_B \} = 0$,
\begin{align}\label{}
\{\nabla^-_A , \pa^+_B\} =  W_{AB}(x,\q,u)\,.
\end{align}
It corresponds to the harmonic projection with $u^-_\a u^+_\b$ of the first relation in \p{constr}.
The absence of a term symmetric in $A,B$ on the right-hand side  is the defining constraint of the theory. 
From here we derive $W_{AB}= \pa^+_B A^-_A$. We have already found $A^-_{A}$, eq.~\p{3.9}, so we obtain \cite{Chicherin:2016fac}
\begin{align}\label{Wab}
W_{AB}(x,\q,u) = \pa^+_{A} \pa^+_{B} A^{--}\,.
\end{align}
The expression on the right-hand side is manifestly antisymmetric in $A,B$, so the constraint has been solved. 
Thus, the chiral $W_{AB}$ is obtained from the harmonic gauge connection $A^{--}$ by acting with two odd harmonic-projected derivatives.
We see that the gauge connection $A^{--}$ for the harmonic   derivative $\pa^{--}$ is the {\it only object we need for constructing  chiral 
supercurvatures and  composite operators}.  According to \p{448}, $A^{--}$ is itself expressed in terms of the L-analytic prepotential $A^{++}$, and we can rewrite eq.~\p{Wab} directly in its terms,
\begin{align}\label{WabFull}
\boxed{
W_{AB}(x,\q,u) = \pa^+_A \pa^+_B \sum^\infty_{n=1} (-1)^n\int du_1\ldots du_n\; { 
A^{++}(x,\q \cdot u^+_1,u_1) \ldots A^{++}(x,\q \cdot u^+_n,u_n) \over (u^+u^+_1)(u^+_1u^+_2) \ldots (u^+_nu^+)}}
\end{align} 

Our construction guarantees the covariance of the supercurvature 
under the gauge transformations \p{21} with an L-analytic parameter $\Lambda$,
\begin{align} \label{Wcov}
\delta_{\Lambda} W_{AB} = [W_{AB},\Lambda]\,.
\end{align}
Let us check this using the explicit expression \p{Wab}. 
The gauge variation $\delta_{\Lambda}W_{AB}$ is expressed 
through $\delta_{\Lambda} A^{--} = \nabla^{--} \Lambda$, eq.~\p{21}. Then we apply the obvious corollary of the 
(anti)commutation relations $[\pa^{--}, \pa^+_A] = \pa^-_A$ and $\{\pa^+_A, \pa^-_B\}=0$:
\begin{align} \label{lem}
\text{If} \ \ \Phi=\Phi(x,\q^+,u) \ \ \text{is L-analytic} \ \ (\pa^{+}_A \Phi = 0)
 \ \ \Rightarrow \ \ \pa^+_A \pa^+_B \pa^{--} \Phi = 0 \,,
\end{align}
with $\Phi \to \Lambda$ and we obtain \p{Wcov}.

The supercurvature carries $U(1)$ charge zero 
and gives rise to an infinite expansion \p{3} in the harmonic variables. 
However, it is (covariantly) harmonic independent in the sense that $\nabla^{++} W_{AB} = 0$ (recall \p{Han}).
Indeed, the defining equation for the gauge connection $A^{--}$, $[\nabla^{++},\nabla^{--}] = \pa^0$ (recall \p{covar}) is
equivalent to $\nabla^{++} A^{--} = \pa^{--} A^{++}$. Using \p{lem} with $\Phi \to A^{++}$ we get
\begin{align}\label{314}
\nabla^{++} W_{AB}  = \pa^+_A \pa^+_B \nabla^{++} A^{--}= \pa^+_A \pa^+_B \pa^{--} A^{++}=0\,.
\end{align} 
This implies that the dependence on the harmonics can be eliminated  by going to another gauge frame (see the next section), although this step becomes unnecessary for composite gauge invariant operators constructed out of $W_{AB}(x,\q,u)$. 
For example, consider the chiral Konishi multiplet defined in \p{K}. It is easy to see that in fact it does not depend on the harmonic variables. Indeed, we act on $K(x,\q,u)$ with the 
harmonic derivative $\pa^{++}$, eq.~\p{4}, which is replaced by the covariantized harmonic derivative under the trace. Then \p{314} implies
\begin{align}\label{Kchir}
\pa^{++} K(x,\q,u)  = \frac{1}{2}\ep^{ABCD} \tr \,\nabla^{++} (W_{AB} W_{CD}) = 0\,.
\end{align} 
 Since $K$ carries $U(1)$ charge zero, the property \p{Han} yields its independence of the harmonics, $K(x,\q,u) = K(x,\q)$. We thus see that although the supercurvature $W_{AB}$ depends on the harmonics, the gauge invariant operators formed from it are harmonic independent.
If one prefers to see the harmonic independence manifestly, one can integrate $K(x,\q,u)$ on $S^2$, $K(x,\q) = \int du\, K(x,\q,u)$ according 
to eq.~\p{6}.

The operators we consider are gauge invariant, so we have the right to substitute in \p{Wab} the 
expression for $A^{--}$ in the Wess-Zumino gauge, eq.~\p{512}. From this we see that 
only the scalars $\phi$, chiral gluinos $\psi^{\a A}$ and the Lagrange multiplier $G_{\a\b}$ contribute to the chiral truncation $K(x,\q)$. What about the composite operators including the anti-chiral gluinos $\bar\psi^{\da}_{A}$ and the YM covariant derivatives $\nabla^{\da\a}$ ? They all live in the non-chiral sector of $K(x,\q,\bq)$ and
are not present in the chiral truncation $K(x,\q)$. In \cite{Chicherin:2016fbj} we explained in detail how to reconstruct the non-chiral sector by means of $\bar Q$-supersymmetry transformations. It is in this sector that the second prepotential $A^+_\da$ starts playing its role.

This concludes our summary of the construction of gauge invariant composite operators first presented in  \cite{Chicherin:2016fac,Chicherin:2016fbj}.

\section{Bridges and frames}

Now we would like to discuss in some more detail the relationship between the Lorentz harmonic and twistor approaches. We show that all the equations from Sect.~\ref{CompOp} are immediately translated into the twistor language, with some small adjustments. 
In particular, in the twistor framework one intensively uses the notion of a `parallel propagator' \cite{Mason:2005zm,Mason:2010yk,Bullimore:2011ni}.
Here we show how to construct it in the harmonic framework and we argue that it is not really necessary, at least for the purpose of constructing composite operators. The arguments presented here follow closely the harmonic interpretation  \cite{Galperin:1987wc} of the Ward construction of instantons, as well as the formulation of $\cN=2$ SYM in harmonic superspace \cite{Galperin:1984av}.

In the preceding Section we worked in the so-called analytic gauge frame,  in which the gauge connections transform with an L-analytic parameter $\Lambda$, eq.~\p{21}. It is possible to switch to the so called $\tau$- (or central) frame \cite{Galperin:1984av,Mason:2005zm}. There the $SU(2)_L$ algebra of harmonic derivatives becomes flat and but $\pa^+_{A}$ acquires a gauge connection instead,
\begin{align}\label{}
\{ \nabla^{++} \ , \ \nabla^{--}\ ,\ \pa^0\ ,\ \pa^+_A\ ,\ \nabla^-_A\ ,\ \nabla^{\pm}_{\da} \}\ \xrightarrow{h}\ 
\{ \pa^{++}\ ,\ \pa^{--}\ ,\  \pa^0\ ,\ \nabla^+_A\  ,\ \nabla^-_A\ ,\ \nabla^\pm_{\da} \}\,.
\end{align}
The `bridge' relating the analytic and $\tau$-frames has the form of a generalized finite gauge transformation  $h(x,\q,u)$  \cite{Galperin:1984av,Galperin:1987wc}.
In particular, $h^{-1} \nabla^{++} h = \pa^{++}$ and $h^{-1} \pa^{+}_{A} h = A^+_A$.
In the $\tau$-frame the gauge transformations of the superfields are $\delta_{\tau} A = \nabla \tau$ with $\nabla = \pa + A$. The parameter $\tau = \tau(x,\q)$ is chiral and does not depend on the harmonics, since the harmonic derivatives are flat.
The bridge $h$ undergoes gauge transformation with respect to both the analytic and $\tau$-frames,
\begin{align} \label{gaugebridge}
h(x,\q,u) \ \rightarrow \  e^{-\Lambda(x,\q^+,u)}h(x,\q,u) e^{\tau(x,\q)}  \,.
\end{align}

In Sect.~\ref{CompOp} we discussed the supercurvature $W_{AB}(x,\q,u)$ \p{WabFull} in the analytic frame. It depends on the harmonic variables, 
but is covariantly harmonic independent, eq.~\p{314}. 
Using the bridge $h$ we can strip off the dependence on the harmonics, thus obtaining the familiar chiral supercurvature $W_{AB}(x,\q)$,
\begin{align} \label{Wnoharm}
W_{AB}(x,\q) = h^{-1}(x,\q,u) W_{AB}(x,\q,u) h(x,\q,u)\,,
\end{align}
which is the truncation of the full supercurvature $W_{AB}(x,\q,\bq)$ in \p{constr} for $\bq \to 0$.
We emphasize once more that the elimination of the harmonics according to eq.~\p{Wnoharm} is an unnecessary step if one is interested in constructing gauge invariant objects out of $W_{AB}$. Indeed, the bridge transformation \p{Wnoharm} obviously drops out from, e.g., the Konishi multiplet \p{K}. 

The bridge $h$ can be found by solving the differential equation $\nabla^{++} h(x,\q,u) = 0$. However, due to the $\tau$-frame gauge freedom \p{gaugebridge}, the solution for $h$ is not unique. 
Nevertheless, it is possible to combine a pair of $h$ bridges into 
the so-called `parallel propagator' $U(x,\q;u,v)$ which depends on two sets of harmonics $u$ and $v$,
\begin{align}\label{Uh}
U(x,\q;u,v) = h(x,\q,u) h(x,\q,v)^{-1}\,.
\end{align} 
In \cite{Bullimore:2011ni} $U(u,v)$ is interpreted as a holomorphic Wilson line. We prefer to call it a bridge between two analytic frames. It is inert under the $\tau$-frame gauge transformations but transforms with respect to both
analytic frames,
\begin{align} \label{gb}
 U(x,\q;u,v) \ \rightarrow \  e^{-\Lambda(x,\q^+,u)} U(x,\q;u,v) e^{\Lambda(x,\q^+,v)}\,.
\end{align} 
The bridge $U$ satisfies the same differential equation on $S^2$ as the bridge $h$,
supplemented with a boundary condition: 
\begin{align}\label{Udiff}
\nabla^{++}_u U(x,\q;u,v) = 0 \ \ , \ \ U(x,\q;u,u) = 1\,.
\end{align}
This makes the solution  unique, given by the formula \cite{Bullimore:2011ni}
\begin{align} \label{U}
U(u,v) = 1 + \sum_{n = 1}^{\infty} (-1)^n \int d u_1 \ldots d u_n
\frac{(u^+ v^+)A^{++}(1)\ldots A^{++}(n)}{(u^+ u_1^+)(u_1^+ u_2^+) \ldots (u_n^+ v^+)} \,,
\end{align}
with $A^{++}(k) \equiv A^{++}(x,\q \cdot u^+_k,u_k)$. Notice that all the local $U(1)$ charges are balanced in \p{U}. The differential equation in \p{Udiff} 
can be easily checked along the lines of the proof in  Sect. 3.3 of \cite{Chicherin:2016fac} 
that $A^{--}$ \p{512} solves its defining equation \p{covarr}. The boundary condition in \p{Udiff} follows from the factor $(u^+ v^+)$ in \p{U}. The bridge $U$ looks very similar to the prepotential $A^{--}$, eq.~\p{448}. Indeed,  the latter is obtained as the limit of the former \cite{Lovelace:2010ev},
\begin{align}
A^{--}(u) = \left. \frac{U(u,v) - 1}{(u^+ v^+)} \right|_{v \to u}  \,.
\end{align}
In the harmonic approach we prefer to deal with the harmonic gauge connection $A^{--}$ which suffices to construct gauge-invariant objects. The bridges $h$ and $U$ are thus superfluous for this purpose.

Finally, using the bridge $U$ we can rewrite eq.~\p{WabFull} in an equivalent form 
by distributing the two odd derivatives on the various prepotentials $A^{++}$ in the expression for $A^{--}$ \p{448},
\begin{align} \label{WabU}
&W_{AB}(x,\q,u) = -\int d v\, \frac{U(u,v)}{(u^+ v^+)}\, \pa^{+}_{u;A} \pa^{+}_{u;B} A^{++}(v) \,\frac{U(v,u)}{(v^+ u^+)} \nt
&+ \int d v d w \, \frac{U(u,v)}{(u^+ v^+)}\, \pa^{+}_{u;A} A^{++}(v)\, \frac{U(v,w)}{(v^+ w^+)}
\, \pa^{+}_{u;B} A^{++}(w)\, \frac{U(w,u)}{(w^+ u^+)} - (A \leftrightarrows B)\,,
\end{align}
where we specify only the harmonic dependence since all the fields are sitting at the same $(x,\q)$ point; 
the lower index $u$ in $\pa^+_{u;A} \equiv u^{+\a}\pa_{\a A}$ specifies the harmonic projection.
The first line in eq.~\p{WabU} has been proposed in \cite{Adamo:2011dq} (in twistor notation) 
as a natural guess for the construction of composite operators. However, such an expression cannot be gauge covariant by itself, as clearly follows from our derivation. One does need two total odd derivatives acting on $A^{--}$, recall \p{lem}. 
The second line in eq.~\p{WabU}  restores the gauge covariance.
One can strip off the dependence on the harmonic $u$ in eq.~\p{WabU} 
according to eq.~\p{Wnoharm}. As a result, the leftmost and rightmost factors of $U$ in \p{WabU} are replaced by bridges $h$, recall \p{Uh}. The bottom component of this expression, 
i.e. $W_{AB}(x,\q =0) = \phi_{AB}(x)$, has been put forward in \cite{Koster:2016ebi}, eq. (12), as a proposal.

We hope that our discussion clearly shows that eq.~\p{WabU} is equivalent to eq.~\p{WabFull}
which appeared earlier in \cite{Chicherin:2016fac}. Moreover, our argument explains the simple geometric origin of the construction of composite operators.


\section*{Acknowledgements}

We are grateful to Lionel Mason for correspondence. 
We acknowledge partial support by the French National Agency for Research (ANR) under contract StrongInt (BLANC-SIMI-4-2011). The work of D.C. has been supported by the ``Investissements d'avenir, Labex ENIGMASS'' and partially supported by the RFBR grant 14-01-00341.



\begin{thebibliography}{99}

 
\bibitem{Chicherin:2016fac}
  D.~Chicherin and E.~Sokatchev,
  ``N=4 super-Yang-Mills in LHC superspace. Part I: Classical and quantum theory,''
  arXiv:1601.06803 [hep-th].
	

\bibitem{Galperin:1984av}
  A.~Galperin, E.~Ivanov, S.~Kalitsyn, V.~Ogievetsky and E.~Sokatchev,
  ``Unconstrained N=2 Matter, Yang-Mills and Supergravity Theories in Harmonic Superspace,''
  Class.\ Quant.\ Grav.\  {\bf 1} (1984) 469
   [Class.\ Quant.\ Grav.\  {\bf 2} (1985) 127].
	

\bibitem{Galperin:2001uw}
  A.~S.~Galperin, E.~A.~Ivanov, V.~I.~Ogievetsky and E.~S.~Sokatchev,
  ``Harmonic superspace,''
  Cambridge, UK: Univ. Pr. (2001) 306 p

\bibitem{Galperin:1987wc}
  A.~S.~Galperin, E.~A.~Ivanov, V.~I.~Ogievetsky and E.~Sokatchev,
  ``Gauge Field Geometry From Complex and Harmonic Analyticities. Kahler and Selfdual {Yang-Mills} Cases,''
  Annals Phys.\  {\bf 185} (1988) 1.


\bibitem{Evans:1992az}
  M.~Evans, F.~Gursey and V.~Ogievetsky,
  ``From 2-D conformal to 4-D selfdual theories: Quaternionic analyticity,''
  Phys.\ Rev.\ D {\bf 47} (1993) 3496
  [hep-th/9207089].
		
		
\bibitem{Lovelace:2010ev}
  C.~Lovelace,
  ``Twistors versus harmonics,''
  arXiv:1006.4289 [hep-th].
		
		
\bibitem{Adamo:2011dq}
  T.~Adamo, M.~Bullimore, L.~Mason and D.~Skinner,
  ``A Proof of the Supersymmetric Correlation Function / Wilson Loop Correspondence,''
  JHEP {\bf 1108} (2011) 076
  [arXiv:1103.4119 [hep-th]].
	
\bibitem{Adamo:2011cd}
  T.~Adamo,
  ``Correlation functions, null polygonal Wilson loops, and local operators,''
  JHEP {\bf 1112} (2011) 006
  [arXiv:1110.3925 [hep-th]].
  
\bibitem{Alday:2010zy}
  L.~F.~Alday, B.~Eden, G.~P.~Korchemsky, J.~Maldacena and E.~Sokatchev,
  ``From correlation functions to Wilson loops,''
  JHEP {\bf 1109} (2011) 123
  [arXiv:1007.3243 [hep-th]].
  
\bibitem{Eden:2010zz}
  B.~Eden, G.~P.~Korchemsky and E.~Sokatchev,
  ``From correlation functions to scattering amplitudes,''
  JHEP {\bf 1112} (2011) 002
  [arXiv:1007.3246 [hep-th]].
	
	
\bibitem{Chicherin:2014uca}
  D.~Chicherin, R.~Doobary, B.~Eden, P.~Heslop, G.~P.~Korchemsky, L.~Mason and E.~Sokatchev,
  ``Correlation functions of the chiral stress-tensor multiplet in $ \mathcal{N}=4 $ SYM,''
  JHEP {\bf 1506} (2015) 198
  [arXiv:1412.8718 [hep-th]].
		
\bibitem{Koster:2016ebi}
  L.~Koster, V.~Mitev, M.~Staudacher and M.~Wilhelm,
  ``Composite Operators in the Twistor Formulation of $\mathcal{N}=4$ SYM Theory,''
  arXiv:1603.04471 [hep-th].

\bibitem{Koster:2014fva}
  L.~Koster, V.~Mitev and M.~Staudacher,
  ``A Twistorial Approach to Integrability in $\mathcal N=$ 4 SYM,''
  Fortsch.\ Phys.\  {\bf 63} (2015) no.2,  142
  [arXiv:1410.6310 [hep-th]].

\bibitem{Mason:2005zm}
  L.~J.~Mason,
  ``Twistor actions for non-self-dual fields: A Derivation of twistor-string theory,''
  JHEP {\bf 0510} (2005) 009
  [hep-th/0507269].
  
\bibitem{Boels:2006ir}
  R.~Boels, L.~J.~Mason and D.~Skinner,
  ``Supersymmetric Gauge Theories in Twistor Space,''
  JHEP {\bf 0702} (2007) 014
  [hep-th/0604040].


\bibitem{Adamo:2011cb}
  T.~Adamo and L.~Mason,
  ``MHV diagrams in twistor space and the twistor action,''
  Phys.\ Rev.\ D {\bf 86} (2012) 065019
  [arXiv:1103.1352 [hep-th]].

			 

\bibitem{Roslyi:1989ya}
  A.~A.~Roslyi,
  ``Superyang-mills Constraints As Integrability Conditions,''
  IN *ZVENIGOROD 1982, PROCEEDINGS, GROUP THEORETICAL METHODS IN PHYSICS, VOL. 3* 587-593.
  
\bibitem{Roslyi:1986yg}
  A.~A.~Rosly,
  ``Gauge Fields in Superspace and Twistors,''
  Class.\ Quant.\ Grav.\  {\bf 2} (1985) 693.

   
 
\bibitem{Devchand:1992st}
  C.~Devchand and V.~Ogievetsky,
  ``Superselfduality as analyticity in harmonic superspace,''
  Phys.\ Lett.\ B {\bf 297} (1992) 93
  [hep-th/9209120].
 
  
\bibitem{Devchand:1993ba}
  C.~Devchand and V.~Ogievetsky,
  ``The Matreoshka of supersymmetric selfdual theories,''
  Nucl.\ Phys.\ B {\bf 414} (1994) 763
  [hep-th/9306163].

\bibitem{Devchand:1995gp}
  C.~Devchand and V.~Ogievetsky,
  ``Selfdual supergravities,''
  Nucl.\ Phys.\ B {\bf 444} (1995) 381
  [hep-th/9501061].
  


\bibitem{Zupnik:1987vm}
  B.~M.~Zupnik,
  ``The Action of the Supersymmetric ${\cal N}=2$ Gauge Theory in Harmonic Superspace,''
  Phys.\ Lett.\ B {\bf 183} (1987) 175.
  
\bibitem{Siegel:1992xp}
  W.~Siegel,
  ``N=2, N=4 string theory is selfdual N=4 Yang-Mills theory,''
  Phys.\ Rev.\ D {\bf 46} (1992) 3235
  [hep-th/9205075].
  
\bibitem{Sokatchev:1995nj}
  E.~Sokatchev,
  ``An Action for N=4 supersymmetric selfdual Yang-Mills theory,''
  Phys.\ Rev.\ D {\bf 53} (1996) 2062
  [hep-th/9509099].

   
   
\bibitem{Witten:2003nn}
  E.~Witten,
  ``Perturbative gauge theory as a string theory in twistor space,''
  Commun.\ Math.\ Phys.\  {\bf 252} (2004) 189
  [hep-th/0312171].
    
  
\bibitem{Chalmers:1997sg}
  G.~Chalmers and W.~Siegel,
  ``Dual formulations of Yang-Mills theory,''
  hep-th/9712191.
			
			
\bibitem{Penrose:1986ca}
  R.~Penrose and W.~Rindler,
  ``Spinors And Space-time. Vol. 2: Spinor And Twistor Methods In Space-time Geometry''
				

 
\bibitem{Sohnius:1978wk}
  M.~F.~Sohnius,
  ``Bianchi Identities for Supersymmetric Gauge Theories,''
  Nucl.\ Phys.\ B {\bf 136} (1978) 461.
					
					
\bibitem{Chicherin:2016fbj}
  D.~Chicherin and E.~Sokatchev,
  ``N=4 super-Yang-Mills in LHC superspace. Part II: Non-chiral correlation functions of the stress-tensor multiplet,''
  arXiv:1601.06804 [hep-th].


\bibitem{Mason:2010yk}
  L.~J.~Mason and D.~Skinner,
  ``The Complete Planar S-matrix of N=4 SYM as a Wilson Loop in Twistor Space,''
  JHEP {\bf 1012} (2010) 018
  [arXiv:1009.2225 [hep-th]].

 
\bibitem{Bullimore:2011ni}
  M.~Bullimore and D.~Skinner,
  ``Holomorphic Linking, Loop Equations and Scattering Amplitudes in Twistor Space,''
  arXiv:1101.1329 [hep-th].

\end{thebibliography}
\end{document}